\pdfoutput=1

\documentclass[manuscript,authorversion,nonacm]{acmart}

\AtBeginDocument{%
  \providecommand\BibTeX{{%
    \normalfont B\kern-0.5em{\scshape i\kern-0.25em b}\kern-0.8em\TeX}}}

\usepackage{textcomp}
\usepackage{color,soul}
\usepackage{float}

\begin{document}

\title{Not Just a Number: A Multidimensional Approach to Ageing in HCI}

\author{Bran Knowles}
\email{b.h.knowles1@lancaster.ac.uk}
\orcid{} 
\affiliation{
  \institution{Lancaster University}
  \city{Lancaster}
  \country{United Kingdom}
}

\author{Jasmine Fledderjohann}
\email{j.fledderjohann@lancaster.ac.uk}
\affiliation{%
  \institution{Lancaster University}
  \city{Lancaster}
  \country{UK}
}

\author{Aneesha Singh}
\email{aneesha.singh@ucl.ac.uk}
\affiliation{%
  \institution{University College London}
  \city{London}
  \country{UK}
}

\author{Richard Harper}
\email{r.harper@lancaster.ac.uk}
\affiliation{%
  \institution{Lancaster University}
  \city{Lancaster}
  \country{UK}
}

\author{Julia McDowell}
\email{julia.mcdowell@lancaster.ac.uk}
\affiliation{%
  \institution{Lancaster University}
  \city{Lancaster}
  \country{UK}
}

\author{Judith Tsouvalis}
\email{j.j.tsouvalis@lancaster.ac.uk}
\affiliation{%
  \institution{Lancaster University}
  \city{Lancaster}
  \country{UK}
}

\author{Alice Ashcroft}
\email{alice.ashcroft@lancaster.ac.uk}
\affiliation{%
  \institution{Lancaster University}
  \city{Lancaster}
  \country{UK}
}

\author{Yvonne Rogers}
\email{y.rogers@ucl.ac.uk}
\affiliation{%
  \institution{University College London}
  \city{London}
  \country{UK}
}

\author{Ewan Soubutts}
\email{e.soubutts@ucl.ac.uk}
\affiliation{%
  \institution{University College London}
  \city{London}
  \country{UK}
}

\author{Andrew Steptoe}
\email{a.steptoe@ucl.ac.uk}
\affiliation{%
  \institution{University College London}
  \city{London}
  \country{UK}
}

\author{Caroline Swarbrick}
\email{c.swarbrick2@lancaster.ac.uk}
\affiliation{%
  \institution{Lancaster University}
  \city{Lancaster}
  \country{UK}
}


\renewcommand{\shortauthors}{Knowles, et al. \copyright January 2025}

\begin{abstract}
The focus on managing  problems that can arise for older adults has meant that extant HCI and Ageing research has not given the concepts of `age' and `ageing' the explicit theoretical attention they deserve. Attending to this gap, we critically examine a ten-year corpus of CHI publications through the lens of an existing typology which we have further developed to analyse how age is understood, interpreted and constructed in the field of HCI. Our  resulting multidimensional typology of age in HCI elucidates the distinctive characteristics of older adults considered when designing with and for this user group, but also highlights the need for a more critical, reflexive, social constructivist approach to age in HCI. Applying this approach, we explore age as a multidimensional system of stratification to better understand the phenomenon of the age-based digital divide. 

\end{abstract}



\maketitle

\section{Introduction}

An interest in people in later life has been evident in HCI for several decades \cite{vines2015age}, with dedicated
subcommittees on Health, Accessibility and Aging (2017–2018) and Accessibility and Aging (2019–present)
emerging at the field's flagship conference, CHI.  Within this literature are recurring admonitions to avoid homogenising older adults \cite{durick2013dispelling,lazar2017going,Light16,petrie2023talking} and, relatedly, to challenge negative attitudes towards ageing (e.g.\,as a process of physical and cognitive decline) \cite{knowles2021harm,soro2017minding,vines2015age,DHaeseleer19}. This has led to a significant expansion of the subfield's remit and a more multifaceted interest in the diverse needs, desires, activities, and attitudes of older adults. This has advanced the field in many ways while also raising fundamental questions around what might unify or organise how the CHI community approaches ageing---specifically, \textit{who} would this community consider to be an `older adult',  what is it about \textit{age} that matters to HCI, and (how) does this show itself in terms of how interaction with computers is designed? 

The answers to these questions are consequential beyond the mundane activities of a research subfield constituting itself. They reflect, while also contributing to, how age is, in important respects, 
 \textit{socially constructed}. There are measurable physiological changes associated with ageing, but what age means to those categorised as such is in itself a social fact, its ramifications perhaps more consequential than some aspects of physiology. Despite HCI developing more nuanced discourses on other categories that are socially constructed, such as race and gender \cite{schlesinger2017intersectional,ashcroft2023reflexivity,de2019feminist,erete2023method,rankin2019straighten}, age and ageing have not been commonly explored or even acknowledged within HCI research as similarly social. In this social constructionist view two points should be clear:  First, age categories used to distinguish groups of people vary in accordance with the specific meanings, values, and symbols imbued by different cultures;  and second,  ageing is not a wholly biologically determined and chronologically measurable process, but also a \textit{social process}. 

    In this paper we explore the implications of these two interrelated points for the way HCI research approaches the core concepts of `age' and `ageing'---first in terms of how `problems' (seemingly) stemming from age are understood and addressed; and, second, in terms of larger themes within the literature, specifically the age-based digital divide, which can be seen as having its own rhetorical properties connected to the social construction of this posited phenomenon (the `digital divide'). This does not invalidate the notion of the digital divide, but might force more demanding research into what it might be.   

    To support these claims, we adapt Johfre and Saperstein's \cite{johfre2023social} synthesis of the dimensions of age examined across multi-disciplinary age research.     Our version of their typology allows us to systematically document the myriad ways age has been conceptualised 
    within  a ten-year corpus of extant CHI literature. Through this process, we make several important contributions to the HCI literature on ageing (hereafter: `HCI and Ageing'): 
\begin{itemize}
\item  First, through extending the social construction of age framework to the CHI literature, we highlight how this research and advancement of digital technology are implicated in the ever-evolving social construction of age and ageing;
    \item Second, we illustrate this by addressing a core ambiguity in the HCI and Ageing field about what  is important to HCI about being `older', showing without careful reflection on social factors or institutional processes and structuring, there is  potential for research to construct ageing as naturally begetting exclusion through declining ability;  
    \item But we also show, third of all, that a systematic coding of extant literature against an adapted typology uncovers an impressively extensive coverage of measures (and concerns arising in relation to these measures) within HCI that could negate this reflexive narrowing; 
    \item Fourth, by adapting Johfre and Saperstein's typology, we develop a much needed framework to organise the profundities that motivate this community, offering a holistic view of how multiple dimensions of age intersect in HCI and why this matters; 
    \item And finally, we extend beyond Johfre and Saperstein by demonstrating how age operates as a multidimensional system of stratification\footnote{Johfre and Saperstein use `system of inequality’ rather than `system of stratification’. While `inequality' is suggestive of an outcome, `stratification' usefully draws attention both to the structural process of constructing a hierarchy and the outcome. This linguistic shift enables us to differentiate between inequalities that reflect the agency of individual actors making choices compared to the inequitable structural process of creating a social hierarchy.} that contributes to the digital divide and associated notions in very particular ways, and how the negative consequences of this might be ameliorated by more insightful, `constructivist' points of view in HCI and Ageing research. 
\end{itemize}

\section{Multiple Dimensions of Age}
\label{framework}

Our argument is that HCI and Ageing research has not made sufficient use of the insight, well-established in the social sciences, that age is (in part) socially constructed. 
That age is socially constructed is evident in the varied ways age manifests itself as socially significant---in how people interpret age cues of others to inform their interactions, in how expectations are formed of the appropriate behaviours and allowances for people of a given age, and in how age is performed by individuals to shape how they are perceived and treated. People of the same chronological age may be seen as (or, indeed, feel) younger or older depending on a variety of factors and their salience in a given social context. This multidimensionality is the subject of explicit theorisation in subdisciplines of age research. There have been numerous calls to unify these theorisations into a multidimensional framework. While earlier attempts have been made to organise macro-level constructs of age,\footnote{e.g., \citet{riley1994age}, who synthesise disciplinary perspectives on how social structures interact with and socially construct age; \citet{north2013subtyping}, who explore how ageism manifests differently for subtypes of young-old and old-old; and \citet{barrett2022centering}, who identifies three levels at which age operates.} Johfre and Saperstein \cite{johfre2023social} are the first to develop an individual-level multidimensional framework of age. This is particularly significant given, as they note, the vast majority of empirical research examines constructs pertaining to individual age.

To develop this framework, Johfre and Saperstein began with a broad review of conceptualisations of age arising from gerontology, management, and critical age studies, extracting only the individual level dimensions and associated measures. They then combined substantively similar dimensions using different terminology, generating nine distinct age dimensions with related subdimensions. 
Dimensions they identified include chronological, generation, physical, psychological, life stage, responsibility, experience, cultural consumption, and `other signifiers' (see Table 2, p.348-349 of \cite{johfre2023social}, partially recreated in our Tables 1--2). 
Subdimensions are represented in the typology via measures used to explore them within empirical research. Any given subdimension is explored in various ways reflective of the disciplinary norms of the fields covered by their analysis; operationalisation notes summarise this variation. The dimensions were further organised in terms of their conceptual closeness to chronological age, from top to bottom of the typology, rather than, e.g., by how commonly these dimensions appear in the literature.  Although they note the typology is not exhaustive, it covers the most well-known and widely used ways of conceptualising and measuring age in social science research. Johfre and Saperstein explore, from their perspective as sociologists, the utility of a multidimensional approach as a framework for examining the complexities that give rise to inequalities.

In \textsection \ref{operationalisation} we adapt Johfre and Saperstein's typology to the context of HCI, narrowing the focus of the operationalisation to the specific concerns of HCI researchers. 
We selected this particular typology for our analysis because HCI typically focuses at the individual level in understanding people's interactions with computers---evidenced by the extensive coverage of individual dimensions in our analysis of the literature. In \textsection \ref{inequality}, we apply Johfre and Saperstein's framework  to explore how these dimensions of age combine to produce a system of stratification reflected in the age-based digital divide.

\section{Methods}
\label{methods}

To systematically catalogue the reasons `older adults' are studied as a distinct category within HCI and what these reasons reveal about how HCI and Ageing conceives of `old age', we conducted a systematic literature review, guided by the standards of the Preferred Reporting Items for Systematic Review and Meta-Analysis (PRISMA) Statement \cite{sarkis2021properly}, detailed in the Appendix. 

\subsection{Corpus}
Our inclusion criteria were CHI Proceedings for the years 2014--2024 focused on older adults (broadly defined). Initially, we searched the database for all publications using the keywords ``older adults'' published within this range. We focused on these dates to cover a sufficient period of development within the field directly preceding and subsequent to Vines et al.'s \cite{vines2015age} seminal work calling for a more nuanced approach to ageing in HCI beyond biomedicalisation.  We excluded papers which were not research papers (i.e.\,descriptions of events, such as courses, workshops, panels, SIGS, town halls, or conference sessions), not CHI Proceedings (some were mis-tagged in the Digital Library), or not focused specifically on older adults. Our initial 804 results were reduced to 724 in a first level screening; a further 490 papers were removed in a second level screening for relevance (see Figure 1). Because the focus was on understanding how authors conceptualise and measure age in this literature, we did not apply inclusion/exclusion criteria based on chronological age of research participants; rather, we included any paper that invoked the category of `older adult' 
as a focus of the research. Papers were excluded for lack of relevance, therefore, if they did not centre `older adults', `age', or `ageing' in their enquiries. This was the case for 490 papers which did not directly engage with older adults as a group---such as, for example, those which focused on other stakeholders who interact with older adults rather than the older adults themselves; treated older adults as a test group for future research; referred to older adults in passing, e.g.\,as one of many potential beneficiaries of a technology; and other such indirect engagement with this group (see Appendix for fuller details). This screening resulted in a corpus of 234 publications, which we call Corpus A.

We assumed searching for ``older adults'' anywhere in the text would capture publications adopting alternative phrasings such as ``older people'', ``senior citizens'' and ``elderly'' given it should pick up ``older adults'' within the titles in the reference list. We tested this assumption using a modified approach: Whereas search A looked for ``older adults'' anywhere in the text, search B looked for ``older adults'' and any of the above synonyms, but only searching within the abstract. This returned 222 publications, which were reduced to 199 following the same screenings (Corpus B).

\begin{figure}
  \includegraphics[width=.98\textwidth]{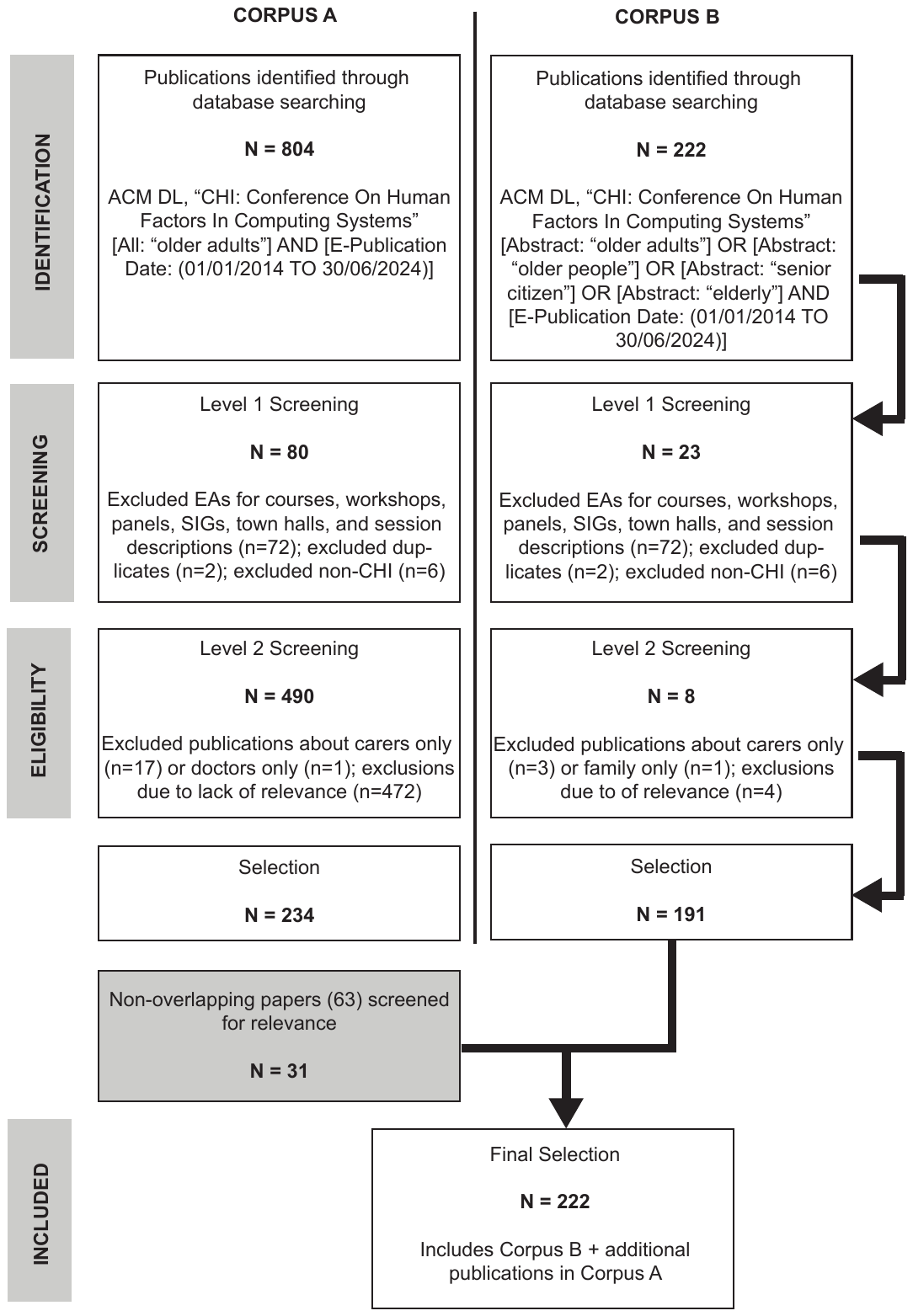}
  \caption{Schematic of corpus creation.}
  \label{fig:PRISMA}
\end{figure}

Our other reason for experimenting with the protocol is how difficult it proved to determine eligibility for many of the papers using method A. We fretted over whether a paper reporting a singular (and perhaps accidental) finding about older adults should be included in a corpus of papers focusing on `older adults'; and also whether papers about physical or cognitive disease ought to be included. Many papers on dementia, for example, explicitly distanced themselves from research about older adults, focusing instead on individuals with dementia at any age. Several publications in Corpus A about dementia were included because they acknowledged it as a disease more prevalent amongst older adults, and yet they did not always seem to satisfy our overarching criteria of exploring older adults as a category for HCI---in other words, older adults comprised a large proportion of samples, but age itself was not the focus of the research. (We note as an aside that this ambiguity is illustrative of the very issues we attempt to address with this paper.) We resolved the matter by re-examining the 63 publications captured by method A but not B: Having coded the abstracts and introductions for Corpus B, we decided to include all papers from Corpus A that would create a code based on the abstract and introduction alone, i.e.\,that mentioned challenges, opportunities, behaviours and/or attitudes of older adults as motivating the research. Thirty-one publications meeting this criterion were added to Corpus B. The final corpus includes 222 publications (see Figure 1).

\subsection{Analysis}
We drew on Johfre and Saperstein's \cite{johfre2023social} multidimensional framework of age for our analysis of this corpus, borrowing the organising structure of  \textit{dimension}, \textit{example measures}, and \textit{operationalisation notes}.  
We adapted it to specifically reflect common measures and conceptual questions in HCI and Ageing (see columns 3 and 4 in Tables 1--2 for a comparison of the original operationalisation with our adaptations to render the typology useful for HCI research). Based on existing knowledge of the HCI and Ageing literature and familiarity with the included papers gained during corpus creation, the lead author created initial operationalisation notes for each measure. These notes were iteratively refined to reflect what was found in the literature until reaching saturation (i.e.\,no new operationalisations were found). At this point, the operationalisation notes were fixed as they appear in column 4 of Tables 1--2, and the whole corpus was re-reviewed using these notes as a guide to ensure coding consistency. During coding, title, abstract, and introduction for all papers were analysed for operationalisations, and the methods section was read in full for any additional information regarding how authors categorised individuals as `older adults' (e.g.\,age range used). Whenever a code was created, exact language used in the papers was captured in the spreadsheet to retain a record of what prompted the code, and any uncertainties were highlighted for additional scrutiny.\footnote{Sometimes sampling techniques were not described in sufficient detail to understand how age was categorised in terms of sampling frame. These cases were flagged for discussion and coding was agreed between the first and second authors.} Publications received as many codes as there were instances of operationalisations. Codes were then moderated by the second author: specifically, all codes were reviewed for a sample of papers (the square root of the total number of papers, n=15),  individual highlighted (uncertain) codes were reviewed, and any differences were discussed and resolved, resulting in minimal revisions of the codes.

\subsection{Limitations}
\label{Limitations} 

We recognise that CHI is not the only relevant venue for HCI and Ageing research. HCI consists of many different constituencies or communities reflecting somewhat different underlying ethoses, methodologies, and possibly, corresponding to this, different constructions of age. Given this,  a similar exercise can and should be undertaken in exploring other settings in which HCI research expresses itself.  CHI is but one, albeit a very large one, as the scale of literature corpus testifies. We acknowledge emphatically that studies of ageing published in, e.g., CSCW, DIS, and ASSETS could produce different results---certainly in terms of the prevalence of various measures (as summarised in Tables 3--4 in the Appendix), but also in terms of the particularities of how measures are operationalised (as narratively summarised in \textsection \ref{operationalisation}). However, while the \textit{social construction} of age may vary across these venues, our detailed, iterative engagement with the sizeable CHI corpus, notable for both its breadth and depth in HCI and Ageing research, suggests the emergence of conceptualisations which could not fit within the typology we present would be unlikely. And while it is outside the scope of this paper to analyse multiple venues---again, given the scale of the CHI literature on this topic, alone---the typology we present represents an invaluable tool for systematically examining characteristic differences between communities.

Our analysis offers a snapshot on a discourse within a conference over a reasonably long sample timeframe.\footnote{We chose not to include TOCHI papers, despite them often being presented at the CHI conference, largely as a practical decision to limit the size of the corpus, but also because TOCHI may differ in subtle ways to the CHI conference which we would not have scope to discuss here.}  While we made every effort to identify and include all publications meeting our criteria, it is possible some are missing, though it is unlikely that we have entirely missed categories pertinent to our typology in Tables 1--2. We include a supplemental bibliography for peer scrutiny of our final corpus and as an added resource for the community. 

Our coding was limited to the title, abstract, introduction, and methods sections of our selected publications. While it is possible that further codes would have been generated if we read papers in their entirety, our approach assumes that if authors' conceptualisation had radically changed in writing the paper it would have been reflected in the abstract or introduction. We also recognise the subjective nature of this type of analysis, and acknowledge  others might have coded differently, though we attempted maximise reliability through code moderation. Moreover, as our aim is to prompt greater attention to how HCI is socially constructing age, the possibility of subjectivity is itself fruitful for prompting debate and discussion, fostering deeper engagement in the field with the consequences of how age is conceptualised and measured.

\section{(De)constructing Age in HCI}
\label{operationalisation}

\subsection{An HCI and Ageing Typology of Age}

In this section, we introduce our adaptation of the Johfre and Saperstein typology with HCI-specific operationalisations. We draw attention to key papers in the corpus to illustrate our development of the typology, as well as its growing utility. (Not every paper in the corpus is discussed below; a full accounting of included papers is provided in the supplementary bibliography.) For ease of comparison, percentages of the measures used to allocate papers into groups in the typology are reported in Tables 3--4 in the Appendix. These percentages provide insight into which measures have received more or less attention in CHI in the past decade, but may not reflect the distribution of interests across the wider HCI and Ageing community; nor do they necessarily reflect beliefs (by those publishing at CHI specifically, nor HCI and Ageing researchers more broadly) about the relative importance of these measures to the matter of `older adults' interacting with computers.

As we will detail below, the CHI literature operationalises these dimensions and measures in a variety of ways. Emulating Johfre and Saperstein, we use our operationalisation notes to summarise these differences, but expand on the variety of operationalisations through narrative summary. Within our description of the dimensions and example measures below, it is important to note the crucial distinction between measures used for the purpose of categorisation and measures which motivate the study of those categorised as `older'. In our analysis, we find chronological age measured as \textit{time since birth} is treated as a first-order sorting criteria for who the researchers consider to be an `older adult'---except, perhaps, for those few publications that do not explicitly state the age range of those they are categorising as `older adults'. (In some instances, \textit{cohort} is used for more precise sorting based on chronological age.) Second-order sorting criteria is used to determine study eligibility if researchers are interested in specific characteristics (e.g.\,particular disabilities, retired, grandparents); but these criteria arise from the motivation to study older adults and are not treated as determinative of old age. In other words, authors typically define `older adults' in terms of a chronological age cut-off (e.g.\,65+), then within that broad categorisation, provide a justification for studying this age group that nods to the other dimensions in the typology (e.g.\,certain physical changes which affect interaction are especially common above age 65). In this sense, the way CHI operationalises these measures tells a story of why CHI studies `older adults' as a special category of user; though, as we explain in more detail in \textsection \ref{takeaways} and \textsection \ref{inequality}, that story in turn influences cultural notions of who fits the category of `older adult', and why that categorisation is useful or meaningful (e.g.\,a person is `old' when they struggle with technology in these particular ways). 

\begin{table*}[p]
    \begin{tabular}{|p{.12\linewidth}|p{.13\linewidth}|p{.25\linewidth}|p{.4\linewidth}|}
        \hline
   \textbf{Dimension}    & \textbf{Example \hspace{20mm} measures} & \textbf{Operationalisation notes}   & \textbf{Operationalisation in HCI \& Ageing} \\ \hline \hline
        Chronological                                                                                           & Time since \hspace{20mm} birth  & Calculated via birthdate (day/month/year, or just year of birth) or asked directly (``How old are you?'')  & Calculated via birthdate/year of birth, or self-reported age of participants
        \\ \hline
         {}                                                                                           & Life expectancy at age X 
  & Calculated based on birth year/cohort; can be subset by health markers, demographics, etc  & Years left to live and use technologies, calculated based on birth year/cohort relative to shifting demographics
  \\ \hline
           {}                                                                                           & Healthy life expectancy at age X 
  & Expected years left to live as healthy and nondisabled & Ability to age healthily, independently, and in one's own home 
  \\ \hline
             Generation                                                                                           & Generation/\hspace{20mm} cohort
  & Birth cohort calculated via chronological age or generational identification (Generation Z, millenial, Generation X, baby boomer, etc.)  & Birth cohort calculated via chronological age bands, shared socio-historical background, or shared exposure to technologies; relative inter-family position \\ \hline 
               {}                                                                                                                                                                                 & Institutional \hspace{20mm}cohort 
  &  Category within age-graded institution, e.g., grade in school & Category within society entitled to social protection and/or requiring management, e.g., vulnerable, needing care;  category within digital economy, e.g., historically marginalised 
  \\ \hline
                 Physical                                                                                          & Biomarkers
  &  Epigenetic modification (e.g., methylation, epigenetic clocks), telomere length, physiological markers & Gait/stride length, skin elasticity, pupil dilation   \\ \hline
                   {}                                                                                          & Health and \hspace{20mm}energy
  &  Sometimes called functional age; includes level of fitness or stamina & Functional abilities (e.g., eyesight, hearing, motor skills, haptic control, balance, swallowing, sensory perception); overall health (e.g., disease prevalence, severity, or co-morbidity); muscle strength and fitness; fatigability \\ \hline
                   {}                                                                                          & Appearance and \hspace{20mm} embodiment
  &  Perceived age, ``look age,'' or separate measures related to body size, sexual maturity, wrinkles, hair color, voice quality, etc. & Voice quality, ``voice age''; speech features; movement repertoire and slowness \\ \hline
                     Psychological                                                                                          & Cognitive \hspace{20mm} ability
  & Performance on a memory task, reaction time, or self-rated  & Processing speed, spatial cognition, cognitive map decay; memory (working, procedural, long-term), input stumble; intelligence (fluid and crystallised) and functional neuroplasticity; social functioning and communication; decision making, multitasking, attentional control/splitting, cognitive inhibition; neurocognitive disorder (mild cognitive impairment and dementia) \\ \hline
                       {}                                                                                          &  Emotional \hspace{20mm} maturity
  & Generally measured through an index that includes aspects such as impulsiveness, indecision, egocentrism, and distractibility  & Measured in rates of technology adoption and interaction behaviours: confidence/hesitance; fearfulness; circumspection; autonomy, collaborativeness; creativity; adventurousness; willingness to learn  \\ \hline
    \end{tabular}
    \caption{\label{tab:table-name1} HCI and Ageing typology of age. Columns 1--3 copied directly from Johfre and Saperstein \cite{johfre2023social}; column 4 describes how the CHI literature on HCI and Ageing operationalises these dimensions of ageing for `older adults'.}
\end{table*}

\begin{table*}[t]
    \begin{tabular}{|p{.12\linewidth}|p{.13\linewidth}|p{.25\linewidth}|p{.4\linewidth}|}
        \hline
   \textbf{Dimension}    & \textbf{Example \hspace{20mm}measures} & \textbf{Operationalisation notes}   & \textbf{Operationalisation in HCI \& Ageing} \\ \hline  \hline
        Life stage                                                                                          &  Age category
  & Membership in nominal categories, e.g., child, adolescent, young adult, middle adult, young-old adult, old-old adult; when self-assessed, sometimes called ``identity age''  & ``Identity age'' (fluid and actively negotiated, e.g., situated elderliness); negative self-stereotyping (self-beliefs about ICT skills); negative other-stereotyping (distancing oneself from `older adult' category)    
  \\ \hline
  {}                                                                 &  Stage in life \hspace{20mm} transitions
  &  Stages could include marriage, homeowner, or retirements status & Stages include retirement (financial stability), social isolation (living alone, distanced family), restricted mobility / disability, (pseudo-) institutionalisation
  \\ \hline
  Responsibility                                                                 &  Family role
  & Social role as parent or child, level of household responsibilities  & Social role as parent, grandparent, spouse; role as carer or cared-for individual; role within intra-family/household technology relations (receiver of IT support, sharer of data/passwords)
  \\ \hline
      {}                                                                 &  Civic rights \hspace{20mm} and responsibilities
  & Whether the person can drive, buy alcohol, or use Medicare  & Whether the person can still drive and contribute to healthcare decisions; contributing to society and community (e.g., inter-generational mentorship, co-design, content creation) 
  \\ \hline
        {}                                                                 &  Workplace \hspace{30mm} responsibility
  &  Job title, whether the person is in a managerial role & Whether the person is still in/seeking employment (e.g., need of technology upskilling) 
  \\ \hline
  
          Experience                                                                 &  Tenure
  & Time spent in an institutional context  & Time spent using technologies (e.g., when they first learned); time spent using newer technologies (e.g., familiarity) 
  \\ \hline
            {}                                                                 &  Expertise
  & Expertise in or length of experience  & Technology expertise (e.g., comfort levels using technology, susceptibility to digital harms)
  \\ \hline
              {}                                                                 &  Wisdom
  & Competence or attitudes  & Competence and self-efficacy; attitudes and integrity (e.g., asserting one's values in adopting or rejecting technology)
  \\ \hline
                Cultural \hspace{10mm} consumption                                                                 &  Aesthetic \hspace{20mm} preferences
  &  Preferences in clothing, music, slang, etc.  & Preferences in technology adoption and use; preferences in robot/chatbot/VA appearance and behaviour, icon style, or tacility    
  \\ \hline 

                   
                  {}                                                                 &  Activities and interests
  & Behaviors, orientations, and hobbies (e.g., frequency of sex)  & Behaviours (e.g., frequency of exercise), orientations (what is considered meaningful and worth doing), and hobbies (e.g., crafting/making, gaming, crowdworking, making music, volunteering, lifelong learning)
  \\ \hline
Other \hspace{10mm} signifiers                                                                 &  First names
  &  Age connotation of legal name or nickname  & Age appearance of avatar
  \\ \hline
    \end{tabular}
    \caption{\label{tab:table-name1} HCI and Ageing typology of age (continued).}
\end{table*}

\subsubsection{Chronological}

The measure of \textit{time since birth} was widely used to categorise individuals as `older adults', with 84.2\% of publications providing a chronology-based definition of older adults or lower-bound (37.1\%) and/or providing an age range for their participants (75.1\%). 
Four publications that did not include a definition or age range were using non-`older adult' participants, such as care staff. Of those that provided a definition, the majority (74.4\%) used either 60 (25.9\%) or 65 (49.4\%) as their lower bound. Six (7.4\%) used a lower bound of 50, 12 (14.8\%) used 55. Ages were provided for study participants in 166 publications (75.1\%); a further 16 providing only the mean age of participants, and a further 4 only providing a lower bound for participants. Where mean age of participants was provided, the lowest mean was 60.65, the highest was 86, and the average was age 72.8. Publications were also coded for how many increments of 5 years were included in the study if reporting participant ages,\footnote{Where ranges were not in multiples of 5 years, they were rounded up. E.g.\,a 6-year range = 2 increments because it is more than 1 and less than 3 increments. For this reason, ranges are approximate, e.g.\,3 increments = $\sim$15 years.} or encapsulated by the definition (e.g.\,65+) if no age range was provided. A small number of publications (3) were n=1, so had an increment count of 0. Zero publications included only a single increment ($\sim$5-year span), and 9 included only 2 increments ($\sim$10-year span). Publications most commonly included 4 increments, i.e.\,$\sim$20-year span (n=112, 50.7\%), and 73 papers (33.0\%) had more than 6 increments (>30-year span), including 5 publications with a massive 9-increment span ($\sim$45 years) and 57 with no upper bound to their definition of `older adult'. Where publications provided both a definition of `older adults' and the age range of participants, we note occasional inconsistency between conceptual and pragmatic definitions: 10 included participants below the stated cut-off, only 2 of which can be explained because the study also included `middle-age' people. We also note a small number of publications use the term `elderly' to refer to individuals as young as 60 \cite{Matjeka20,Motahar22} or 65 \cite{Niksirat17}, despite connotations of late old age.

The dimension of chronology also included related or derived concerns for HCI,  
such as in those papers that  operationalised \textit{life expectancy} and \textit{healthy life expectancy at age X}. Research motivated by the former explores how older adults' perception of diminishing time ``left on earth'' (i.e.\,given their age) relates to their tendency to be more discerning in their technology experiences and interactions and to eschew more ``novel'' technologies \cite{Loup17}. Yet 23 papers (roughly 10\%) mentioned demographic ageing as part of their motivation for researching older adults, which clearly not only has implications for life expectancy for those we might categorise as `older adults' but also represents a changing market for technology solutions for different chronologically defined age groups and their associated expectations. 

The related measure of \textit{healthy life expectancy at age X} was a much more prevalent concern in the literature (35.1\%). Especially prominent was the motivation to reduce the likelihood of age-related disease or disability, e.g.\,through promotion of physical and/or cognitive exercise \cite{Du24,Li24designing,Bradwell24,Yang20,Qian24,Xu16,Xu21,Matjeka20,Karaosmanoglu22,Huang22}, and through home/personal health monitoring \cite{Adib15,Hsu17,Wang20social}\footnote{The corpus also included critiques of monitoring technologies \cite{czech2023independence,soubutts2023shifting} and the development of privacy-preserving approaches    \cite{Mujirishvili24,McNeill17}.}  and fall risk detection \cite{uzor2014investigating,wu2018understanding,Garcia14,Sas17}. Other work aimed to extend independence, e.g.\,with assistive technologies \cite{Stegner23,Williams21,Jelen19,Salai21,Williams21} and mobility support tools \cite{Montuwy18,Meurer14,Stein17mobility}, and  better understanding older adults' use of ICT for health/disease self-management \cite{Visser14} and general information seeking \cite{harrington2022s,Desai23,Trajkova20,Karkera23}. Noting the detrimental impact of loneliness on physical wellbeing \cite{bhowmick2023exploring} and the prevalence of accessibility \cite{Shinde24}, mobility \cite{Felberbaum18}, and communication \cite{Welsh18,Lei24} challenges that impact one's ability to socialise, the literature also explores the potential of technologies to promote socialisation and interpersonal connection, e.g.\,through VR environments \cite{Carrasco17,Xu23designing}, communication prototypes \cite{Brereton15}, and shared reminiscence aids  \cite{Kang21}. The totality of these works appear to reflect the cultural hegemony of decline ideology \cite{gullette2018against} and a ``projected fear of old age'' \cite{higgs2020ideology}. There was, however,  notable pushback against this view in research focusing on the ways older adults flourish, as evidenced by their appropriation of technologies to meaningfully engage in life \cite{brewer2016tell,Lazar21makerspace,Huang24,Mao16,Jin24exploring,Carucci19making,thach2023key}. While the majority of works operationalising this measure align with the normative goal of ``successful aging'' \cite{rowe1997successful}, we found one instance of explicitly motivating the research with a concern about HCI's role in promoting ``positive aging'' \cite{blythe2015solutionism}, and others, more implicitly motivated by positive aging, which examine activities older adults find meaningful and fulfilling \cite{Lazar17successful,Zhao24}.

\subsubsection{Generation/cohort}  
This dimension could be expected to relate to age-period-cohort effects, as examined by demographers to understand social change; but with digital technology being so new in human history, we are only starting to have different generations that might experience cohort differences. Perhaps this is why within the HCI corpus, \textit{generation} (n=16, 7.2\%) was typically seen as relational---e.g., indicating family position, with many contributions around supporting inter-generational interaction/communication between geographically distanced grandparents and grandchildren \cite{Wei23,Welsh18,Gutierrez15} or older adults and their adult children \cite{Kang21}. Authors also noted subtle generational differences in social goals \cite{hope14,Kang21}, communication styles/limitations \cite{Welsh18,Wei23}, and tech experience levels \cite{hope14,Lee16Pic,Gutierrez15} to negotiate through design. Three papers emphasised values as distinctive to different generations, creating a source of conflict between generations regarding specific design decisions \cite{Sabie22} and broad technology agendas \cite{Meurer14,barros2021circumspect}. Others emphasised generational differences in life experiences with technologies \cite{petrie2023talking} and learned interaction paradigms \cite{Kim23how,Rodriguez18}, potentially creating challenges in using new technologies which are further compounded by differences in learning styles \cite{Pang21,Jin24empowering}; or contributed key insights into how people's socio-historical contexts shape their technology relations in later life \cite{sun2014being,Tang22community}. In using \textit{cohort} (n=16, 7.2\%), authors focused on specific sub-categories that did invoke age, but as it related to such super-ordinate relational categories such as `generation'---e.g.\,`younger old'  \cite{yu2023history}, `oldest-old' \cite{waycott2013older,neves2015my}, `Fourth age' \cite{Carucci19making}---or chronological decades, e.g.\,people in their 70s, 80s, 90s \cite{thach2023key,Brereton15, Bennett16}. A small number drew attention to design-relevant differences between cohorts within the wider category of `older adults' \cite{Hong17visuallink,Xu16,While24glanceable}, in one case noting the generational differences between `younger seniors' and their even older parents \cite{hope14}; and three papers reported findings in terms of 5-year age bands \cite{Kang21,Ji24exploring,Lei24} to enable this kind of comparison. With the exception of \cite{Pang21}, there was a little reflection on the characteristic differences between older adults today and older adults of the past to understand the changing nature of what it means to be `old'. There was also little exploration of norms/normative shifts from a lifecourse perspective,\footnote{For example, gender norms around tech use are probably different generations for `Baby Boomers' vs.\,`Gen X' vs.\,`Millennials' vs.\,`Gen Z' given where and how people use technology (e.g.\,for very gendered office work for `Baby Boomers' vs.\,very gendered gaming for `Millenials' vs.\,much greater gender equity for `Gen Z' gamers).} i.e.\,how shared norms of a generation across the lifecourse could both strongly shape and be shaped by tech engagement, and whether this was unique to each generation/cohort.

In translating the measure of \textit{institutional cohort} (23.9\%), there appear to be two (implicitly if not explicitly) ``age-graded'' \cite{johfre2023social} institutional contexts relevant to the corpus. The first is the broad institution of social care which protects vulnerable groups from harm (n=39). `Older adults' are considered more vulnerable due to matters of physical and cognitive health which entitles them to social care resources, including technology solutions wherever possible, to (i) ease  labour costs of this care and (ii) detect risk and intervene early. Specific health and welfare vulnerabilities in the literature include heightened risk of social isolation due to protective measures in place during the pandemic for vulnerable groups to shield in place \cite{Nurain21older,Motahar22}; increased risk of traffic death \cite{Malik21}; and increased risk of death during natural disasters \cite{Suhaimi22}.
They are also conceptualised as disproportionately vulnerable to digital harm, requiring research into how to better protect them online \cite{nicholson2019if,Ray19woe,nicholson2021training,oliveira2017dissecting,Akter23}. The second relevant institution is the `digital economy', where older adults are understood as a historically excluded or marginalised stakeholder (n=14). Contributions operationalising this measure in this way either (i) explore accessibility requirements for `older adults' to be able to benefit from technologies \cite{Latulipe15,Kopec21}, including studying the potential of voice assistant technology to reduce barriers to use \cite{upadhyay2023studying,Brewer22ifalexa}; or (ii) advocate inclusive design \cite{sin2021digital} or co-design that includes `older adults' \cite{harrington2019engaging}. As Harrington and colleagues \cite{harrington2022s,Harrington22examining} note, however, older adult marginalisation is greater for some individuals depending on intersectional identity.

\subsubsection{Physical} 

While \textit{biomarkers} offer a useful heuristic for determining a person's age in the real world, it is an infrequent measure in HCI, with only 4 instances in the corpus. Biomarkers salient to HCI include (i) skin elasticity \cite{bhowmick2023exploring}, which can cause issues for biometric authentication; (ii) gait/stride length \cite{Kim22}, as this can affect the accuracy of step counting devices for older people, and can also be a critical health indicator \cite{Hsu17}; and (iii) pupil dilation \cite{Barry22}, another critical health indicator.

Unsurprisingly, as it aligns with HCI's noted tendency to biomedicalise ageing \cite{vines2015age}, \textit{health and energy} was commonly used in the corpus (n=55; 24.8\%). Studies frequently noted potential decline in functional abilities as (i) an accessibility issue (e.g., for declining eyesight \cite{Palivcova20interactive,Tobias16older}, hearing \cite{Hong17visuallink}, motor skills \cite{kuang2023enhancing,neves2015my,Ziman18factors}, haptic control \cite{Palivcova20interactive}, and sensory perception \cite{Alldridge20panda}) and as (ii) something that could be mitigated or rehabilitated through exergames (e.g., for balance \cite{Li24designing,mostajeran2020augmented}, swallowing \cite{Huang22}, joint issues \cite{Zhao18gamified}, and post-stroke recovery \cite{Wang16lights}). Neck fatigue and mobility constraints were noted problems for older users of VR technologies in particular \cite{Thach21guidelines,Wu24toward}. Papers also characterised older adults as suffering from poorer overall health in terms of disease prevalence, severity, or co-morbidity, inspiring disease screening tools \cite{Barry22}, health/medication self-management solutions \cite{Visser14,Lee14realtime,Wolters14minimal}, and even physical support (an exoskeleton) to compensate for muscle weakness  \cite{Jung18potential}.

\textit{Appearance and embodiment} was an uncommon measure within the corpus (n=5; 2.7\%). We found no instances in the literature pertaining to how old a person looks, though we did find an interest in how an older person sounds and moves. A person's ``voice age'' can be inferred based on volume, pitch, and speed of verbalisations when encountering interaction (UX) problems \cite{Fan21older} and other characteristic speech features (as captured in a dataset of voice interface interactions of individuals with and without mild cognitive impairment) \cite{Wolters15CADENCE}. Distinctive speech features are also apparent in older adults' search queries \cite{karanam2016age} and empathic communication in online support communities  \cite{Mao16}. As for movement, Matjeka \cite{Matjeka20} notes the characteristic narrowing of people's movement repertoires as they age.

\subsubsection{Psychological} 

Also unsurprisingly (for the same reasons as \textit{health and energy}), \textit{cognitive ability} was strongly represented within the corpus (n=62, 27.9\%), despite having screened out a large number of papers on dementia. 
Often (n=12) non-specific cognitive decline was mentioned in passing as something more or less understood to happen as a natural part of aging, and from as early as age 50 \cite{Xu21}. Declining working memory can affect perception \cite{Felberbaum18,Blasko14spatial,Colombo23spatial,While24glanceable} and processing \cite{Kim23how,Tobias16older,Palivcova20interactive}, potentially contributing to interactional issues for older adults, such as difficulty locating UI features \cite{Yu24reducing,yu2023history}, input stumble when typing \cite{Hagiya16typing}, and becoming easily overburdened \cite{Hu22} by multitasking \cite{Li24designing} and attention splitting \cite{Felberbaum18,Felberbaum20}, particularly when compounded with  decreased cognitive inhibition \cite{Du24}. Similar issues, along with declining fluid (but not crystallised) intelligence \cite{karanam2016age}, are thought to contribute to difficulty with decision making, e.g., during a crisis \cite{Zhang20}, exacerbating vulnerability; as well as affect older adults' ability to learn to use technologies \cite{Felberbaum18,Desai23}, which may contribute to digital exclusion. The literature explores opportunities for improving cognitive function through stimulating activities \cite{Brewer16why,Niksirat17,Qian24}, 
compensating for decline with augmenting solutions \cite{Bennett24}, and screening for neurocognitive disorders \cite{gordon2019app,Haddad14interface,ding2022talktive,Barry22,Prange21explainable,Urakami21monitoring}. Other publications operationalising this measure consider the technology needs of (i) individuals with diagnosed mild cognitive impairment (MCI) \cite{mentis2019upside} or dementia \cite{siriaraya2014recreating,Bennett16}, e.g., exploring the potential of memory support tools \cite{Bennett24,Ramos16,Chan20biosignal}, prompting physical activity \cite{Eisapour18participatory}, and investigating opportunities for enrichment \cite{thach2023key}; or (ii) their family and carers \cite{An24easytell,mcdonald2021building,zubatiy2021empowering}, particularly given the ways MCI and dementia impact social functioning \cite{Wang20social} and communication \cite{Jin24exploring,Welsh18}. Also noted are methodological challenges of 
co-designing with cognitively impaired people  \cite{Stegner23} and conducting UX evaluations with people experiencing some forms of cognitive decline \cite{Rodriguez18}.

\textit{Emotional maturity} was less thoroughly operationalised in the corpus (n=9, 4.5\%). Apart from confidence in using technology \cite{nicholson2021training}, emotional maturity was not a measure that declined with age so much as evolved. Traits such as hesitance with new things \cite{Dai15tipper,Lee16Pic} and fearfulness of becoming a victim of cybercrime were evidence of characteristic (and arguably appropriate) cautiousness \cite{nicholson2021training} and circumspection \cite{barros2021circumspect}. The literature further highlights the many ways older adults display creativity, initiative, and autonomy in learning and using technology \cite{Qian24,Jin24empowering,Nurain21older}, their collaborativeness in tasks such as crowdsourcing \cite{Seong20designing}, and their adventurousness in life \cite{ambe2019adventures}.

\subsubsection{Life stage} 

While arguably every paper in the corpus operationalised \textit{age category} in focusing on `older adults' (or whichever category label they chose), very few (n=6, 2.7\%) actually explored how older adults themselves relate to their age category as a `life stage'. Those that did found chronologically older adults frequently distance themselves from the negative stereotype they have of the `older adult' life stage  \cite{caldeira2022hope}, and will co-design solutions for (more impaired) ``other older adults'' rather than themselves \cite{pradhan2020understanding}. The literature also notes being aware of decline in oneself can negatively impact an older person's sense of identity \cite{Rudnik24carejournal} and contribute to negative self-stereotyping \cite{Czech20discovering}, inspiring HCI solutions that promote positive identity for those who do identify as `old'. Only two papers challenged the idea of `older adult' as a ``fixed'' \cite{sun2014being} and ``all-encompassing'' \cite{yu2023history} identity in how they motivated their research, noting this identity is negotiated in context with technology objects that make one feel more or less `old' (see ``situated elderliness'', not in corpus \cite{brandt2010communities}). We note across the corpus the lack of detail regarding the language used to recruit participants, making it difficult to discern whether  participants might be engaging in negative stereotyping when determining their own eligibility for the study.

Life stage also produced \textit{stage in life transitions} (31.1\% of the corpus). A number of papers characterise old age as a stage of adjustment to ``changing needs'' \cite{Williams21} and capabilities \cite{Zhao24,Jin24empowering,gerling2015long},\footnote{We note  a critical gerontology view would assert the role of structural factors (i.e., social welfare and employment policies, notably, retirement) in creating ``structured dependency'' in old age \cite{townsend1981structured}, rather than this being a wholly natural change in needs and capabilities.} e.g., no longer being able to drive \cite{Batbold24,Stein17mobility} and more actively managing health, disability, and care \cite{Mujirishvili24,Lee14realtime,Gupta20towards}. Others challenge this negative portrayal, emphasising the ways older adults can thrive and experience profound happiness and self-acceptance in this life stage \cite{blythe2015solutionism,jones2023s,Zuckerman20}. Transition to retirement\footnote{Publications tended to gloss over the complexities of retirement and generally ignored how people might go in and out of various forms of paid or unpaid work in later life. The highly stratified nature of retirement---across racialised categories, socioeconomic status, global geography, gender, and (dis)ability---is also rarely engaged with.} was recognised as perturbing digital \cite{durrant2017transitions} and financial management behaviours \cite{maqbool2018understanding}, impacting a person's financial resources \cite{Stein17mobility,rogers2014never},\footnote{While a shift from income to pension can mean reduced resources, the literature also notes that older adulthood can, for some, be a stage characterised by financial stability \cite{rogers2014never}. Older adults are frequently targeted by scammers because some have accumulated substantial financial assets \cite{nicholson2019if,oliveira2017dissecting}.} radically changing how people spend their time \cite{rogers2014never,Mao16}, and catalysing identity building \cite{brewer2016tell} and romantic love-seeking  \cite{He23}. Generally, older adulthood was characterised as a lonely stage of life \cite{Erel21,thach2023key} (particularly for older women \cite{Motahar22}), a cause for concern due to the adverse impact of loneliness on overall health \cite{Xu23designing,Sas17} (see from wider ageing literature: \cite{holt2022social,national2020social}). Retirement and bereavement can reduce older people's social networks \cite{bhowmick2023exploring,Lei24}, restricted mobility can create barriers to socialising \cite{Zuckerman20,Shinde24,Carrasco17}, family members may live at a distance \cite{Chen24understanding,norval2014s,Mendel22,Sas17,Kang21} (older adults often live alone \cite{Seo18repowering,Yang24}), and older people regularly experience ``feeling invisible, set aside, and unacknowledged by others'' \cite{Zuckerman20}. CHI contributions include loneliness detection \cite{Ji24exploring}, social isolation prevention / network building \cite{Xing24designing,Yang20,Yuan19beyond,baker2021avatar}, facilitating connection within existing social networks \cite{Axtell19photoflow,Chen24understanding,Kang21}, and understanding how older adults use technology to share and connect with others \cite{vaisutis2014invisible,brewer2016tell,norval2014s}. Other relevant transitions are to pseudo-institutionalisation (i.e.\,living alone but under surveillance) \cite{czech2023independence} and  institutionalisation \cite{neves2015my} (i.e.\,living in an aged cared facility). Works explore how older adults navigate a loss of independence and autonomy \cite{Carucci2019making,Rudnik24carejournal}, how they adjust their technology usage and data sharing to a communal setting \cite{Karkera23,Calambur24,Brewer22ifalexa}, and how technology can enrich the lives of those in residential care facilities \cite{Thach21guidelines,Ullal24iterative,carros2020exploring}.

\subsubsection{Responsibility}

It was reasonably common (8.1\%) for papers to operationalise the measure of \textit{family role}. The literature explores the appropriation of technology by older adults within their social roles as grandparent \cite{Chen24understanding,Cerezo19space,Matjeka20,wallbaum2018supporting,Wei23}, parent (to adult children) \cite{Kang21}, and spouse/carer \cite{mcdonald2021building,Li23any}. There was little exploration of how expectations for various social roles are culturally dependent (and, thus, socially constructed); however, the literature provides insight into how one's experience of ageing is co-produced through relationships with other individuals \cite{Gutierrez2016}, particularly when one is the cared-for individual within a family \cite{While24glanceable,Rudnik24carejournal} or spouse dyad \cite{zubatiy2021empowering}, the person is seeking to retain autonomy while being monitored in their homes \cite{Sas17}, or the person is the recipient of tech support from others \cite{Sharifi23cross,Mendel22}. Further contributions pertaining to responsibility and family role include exploring the risks of older adults offloading their personal digital security and privacy responsibilities onto their carers \cite{mentis2019upside}, and older adults' technology responsibilities as collaborator in their own care \cite{soubutts2023shifting}. In fact, the CHI literature exposes being cared for vs.\,caring for another as an important dimension of responsibility that isn't fully captured in Johfre and Saperstein's measures, as the literature includes examples of older adults being cared for by non-family members. While we opted for a straight translation of the existing typology, we would propose calling this measure \textit{caring role} to better reflect the realities of social reproduction.

\textit{Civic rights and responsibilities} was operationalised in 5.9\% of the corpus. Literature explored driving as a right frequently withdrawn from older adults \cite{Meurer14}, and expectations of their responsibility for data-driven self care \cite{While24glanceable} and shared decision making around health decisions \cite{Constantin19why,Hao24advancing}. But more often, the literature explored how older adults can (
be supported to) participate in \cite{Xing24designing} and contribute to society and their communities. They provide mentorship to younger generations \cite{Yuan19beyond} and develop youths' civic engagement \cite{Cerezo19space}; create content online \cite{brewer2016tell}; can invent technology that benefits others \cite{rogers2014never}; 
and while they are routinely disempowered in directing social change \cite{Sabie22}, they can be empowered through technology \cite{Reuter19older} and through co-design \cite{harrington2019engaging} to participate in civic action. 

Very few papers (n=3) operationalised \textit{workplace responsibility}--unsurprising given the tendency to adopt a chronological definition of `older adults' that aligns with work retirement age.\footnote{We note the use of retirement age to define `older adults' reflects a very socio-culturally specific notion of the lifecourse, based predominantly on a privileged group who can afford to have an extended period of retirement in old age. This is not the reality for many at the bottom of various social hierarchies.} One paper considers how to detect and support task completion difficulty for older workers \cite{Belk15cognimouse}, and another investigates use of online materials for lifelong learning to upskill for an evolving job market \cite{Kim23how}; but both of these adopt lower age bounds for `older adults' (55+ and 56+, respectively). One paper, however, explores the motivations of people aged 60--85 learning programming skills \cite{guo2017older}, one of which being to improve their job prospects.

\subsubsection{Experience} 
\label{experience}

Experience was very shallowly engaged with in the corpus, mirroring its underdevelopment in the wider ageing literature. In Johfre and Saperstein's typology, tenure allows them to reflect on how a person's status tends to increase as a function of duration within an institutional context. Our analysis shows a gap in the literature around understanding how a lifetime of being embedded in social institutions such as the family and workplace shape a person's status in ways that change their experience of technology across the lifecourse. That said, if we adopt a view of technology as an institution \cite{pinch2008technology} in which one can have spent more or less time (say, as a user), then 5.9\% of the corpus operationalised that experience in terms of \textit{tenure} and focused on `older adults' as being new to using, and therefore less familiar with, digital technologies. In this view, their tenure was short, their experience therefore minimal. Even when newness to technology was raised, it was framed in terms of older adults being ``novice'' users  \cite{Harrington23trust,Hornung17,Stewart14assisting,Hagiya16typing}, intertwining \textit{tenure} with \textit{expertise} (14.9\%). Expertise would tend to increase with tenure; however, the literature suggests that older adults can be slower to develop expertise, frequently requiring additional, or at least different, learning/tech support \cite{Batbold24,Pang21,Jin24exploring,Tanprasert24,Jin24empowering,Dai15tipper,zhao2015cofaccade}, and designing in ways that build on older adults' existing competencies \cite{Brereton15}. For this reason, several papers explored how older adults learn to use new technologies \cite{Batbold24,Pang21,mahmud2020learning,upadhyay2023studying,Conte18interactive,Sharifi23cross} and 
adjust to new technology environments \cite{Lazar21makerspace,Carucci19making}. The literature also highlights how low familiarity can provoke anxiety when older adults interact with technologies \cite{Conte18interactive,Hornung17}, how older adults lack confidence in their technical skills \cite{An23weisst}, and how they can be put off by expectations of how hard it might be to learn to use a new technology \cite{Bennett16}. Voice-based interfaces were thought to be particularly appropriate for older users due to being relatively intuitive for newcomers \cite{Ziman18factors,kowalski2019older}. Literature that focused on expertise separate from tenure noted that older adults are less likely to have been formally educated in cyber security \cite{Jones19what} and less likely to learn about security and privacy in conversation \cite{Das18breaking}, both contributing to a lack of skills for effectively managing privacy and security \cite{nicholson2019if,latulipe2022unofficial}; and yet, they have significant expertise in other areas that can benefit technology contexts \cite{Nicenboim18connected,Lai24phenoprobe,ambe2020oldy}, and indeed can master technology \cite{rogers2014never,Carucci2019making}.

\textit{Wisdom} was an important measure of an older adult (11.3\%). The literature appraised older adults as having a characteristically well-developed understanding of what matters to them in life, honed over a lifetime of experience \cite{rogers2014never,sun2014being,Lai24phenoprobe}, which informs deliberate (and highly rational \cite{Anaraky21}) decision making about which technologies to adopt or reject \cite{barros2021circumspect,waycott2016not,Trajkova20,Pang21}. The research finds older adults to have a particularly ``nuanced sensibility'' to agency and control  \cite{ambe2019adventures}, autonomy \cite{Conte18interactive}, dignity \cite{Zhang20}, privacy \cite{Hornung17}, and connection \cite{hope14}. This means that (i) they are highly capable of expressing what they do and don't want from a particular technology when asked \cite{Brewer22ifalexa,Zhao24,Wang24redefining,Jin24exploring,Seong20designing}, and (ii) increased adoption will only come from understanding what underlies their attitudes \cite{caldeira2022hope,Allison23bones,Paneels14smiles} and aligning technology to what older adults value \cite{Ziman18factors,Stegner23,waycott2016not,barros2021circumspect}.

\color{black}

\subsubsection{Cultural consumption} 
\label{cultural}

People's consumption choices reflect accumulated cultural beliefs about age-appropriate preferences. These \textit{aesthetic preferences} (9.0\%) can include choices about not/using (and being seen not/using) particular devices and systems. For example, Sun et al.\,\cite{sun2014being} note that not only do individuals enact what they believe to be age-appropriate use of technology, but they also judge their peers if seen as deviating from these norms. Studies highlighted the stigma older adults themselves associate with the use of technology specifically for older adults \cite{caldeira2022hope}, e.g.\,finding that older adults accept smart rings because they are more discrete \cite{Bilius23}; and how older adults can be embarrassed to wear hearing aids \cite{Hong17visuallink}. In addition to consideration of how older adults prefer to be seen, much of the literature explores their own preferences as it relates to technology they adopt and how they use it. Older adults are found to prefer tangible interfaces \cite{Price18feel,Ziat16frontpanel}, simplicity/minimalism \cite{Zhao24} and skeumorphism \cite{Cho15}, and to dislike inauthenticity in robots \cite{Zuckerman20}; preferences regarding anthropomorphic chatbots (and how they are represented) has also been explored preliminarily \cite{Sin19preliminary,Harrington23trust}. The literature would also suggest that older adults have more well-explored communication preferences for: polite and direct speech \cite{Hu22}, for face-to-face interactions \cite{baker2021avatar}, and for rich and frequent interactions with family \cite{hope14,wallbaum2018supporting,Kang21}, affecting the technologies they choose to adopt.

Older adults' \textit{activities and interests} was a significant focus within the corpus (16.7\%), underpinned by a desire to develop technologies that were perceived as supporting meaningful activities \cite{zhao2023older,Chen24understanding,Paneels14smiles,Zhao24}, and thus inspiring tech adoption. Older adults were found to appropriate technology to satisfy ``bio-psycho-social-spiritual needs'' \cite{Madjaroff16}: They engage in physical activity and may adopt exergames for this purpose \cite{Yang24,Xu16}; they are interested in health information \cite{Desai23}, and in tracking (some aspects of) their health \cite{Kim22,Wang24redefining,While24glanceable}. They like cognitively stimulating activities \cite{upadhyay2023studying,guo2017older,Kim23how,Seong20designing,Brewer16why} (e.g.\,crowdwork and programming and lifelong learning), 
and use technology to improve their mood
\cite{upadhyay2023studying}. They enjoy connecting with others \cite{Xing2023,guo2017older,upadhyay2023studying,norval2014s,vaisutis2014invisible}; and they enjoy creative (crafting, music) \cite{Mao16,Richards17exploring,Seo18repowering,Carucci2019making}, fun (e.g.\,games) \cite{gerling2015long,Carrasco17,Wang16lights,Niksirat17}, and expressive activities \cite{brewer2016tell}.

\subsubsection{Other signifiers}
In their typology \cite{johfre2023social}, Johfre and Saperstein include this miscellaneous dimension and list \textit{first name} as an example measure. As they note, a person's first name can age a person, for example if a name is no longer fashionable. While first name isn't relevant to HCI (except, perhaps, if study participants are choosing their own pseudonyms), \textit{avatar choice} is and doesn't otherwise fit within the typology. Only one publication \cite{Wei23} in the corpus explored how older people choose to represent themselves in virtual reality, finding that they preferred avatars that looked younger than themselves. Given the interest in VR technologies for older adults (n=14, 6.3\%), perhaps greater attention should be paid to the implications of the age appearance of avatars being used.\footnote{See \cite{Carrasco17Negotiating} for a good example of this measure operationalised in HCI literature beyond our corpus.}

\subsection{Characterising `older adults'}
\label{takeaways}

Our analysis highlights numerous aspects of age that matter when designing systems. Indeed, the literature shows there are various outcomes predicted (but not determined) by age---including but clearly not limited to physical and/or cognitive capacities---which can be consequential to an older person's interaction with technology. 
We find that although \textit{time since birth} is used to categorise individuals as `older adults', the invoked matter of relevance to HCI pertains to these other measures in the typology. This is suggestive of a potential disconnect between the criteria we use to determine who is in our studies and the issues that we are most interested in as a research community. Whether this is surprising or not, our investigation of the literature and the typology we have used and developed should enable authors to be more specific about which of the different dimensions and measures might be relevant to the chosen technology context, and inform more deliberate recruitment. If time since birth is not so important, what is? And more likely, not which other factor or dimension, but which set of factors. To support this development, an important next step for the field would be to systematically catalogue the measures and dimensions salient to particular interactional contexts and develop theories of \textit{why} they matter.

Our analysis and typology  also underscores the increasing heterogeneity of individuals as they age (see also: \cite{gregor2002designing}). Not only do people progress along these dimensions at different rates,\footnote{People become grandparents at vastly different chronological ages (or not at all); same for transitioning to retirement; declining physical health may feature prominently in one's experience of ageing but not another's; even dementia, typically associated with advanced old age, can begin as early as age 30.} 
but, as Johfre and Saperstein note,  ``People perform their age and perceive other people's ages through the various age dimensions and how they \textit{combine}'' \cite{johfre2023social} (emphasis added). By age dimensions, they mean all those that can play out when age is seen as a multidimensional matter.  In their view, understanding age and ageing entails capturing how multiple factors intersect. From this perspective, i.e.\,a sociological one, the ones that are most important are socio-structural inequities. But what our research enquiries into the HCI and Ageing literature show is that these are just as important in HCI, and are made more complex and intersecting of each other by cognitive and other factors.  While our analysis shows conceptual overlap between a number of measures---e.g.\,\textit{health and energy} and \textit{healthy life expectancy at age X}; \textit{family/caring role},  \textit{healthy life expectancy at age X}, and \textit{institutional cohort}; \textit{tenure} and \textit{expertise}---our analysis also suggests a more thorough examination of interactions between dimensions should be undertaken to a enable richer multidimensional theorisations of age in HCI.

Putting it more critically, we have found that HCI and Ageing research, like the wider ageing literature critiqued by Johfre and Saperstein, focuses on outcomes of age (the problems age presents and solutions to these problems), and has shown little engagement with how the motivations for researching `older adults' reflect and in turn shape (if only to reinforce) cultural notions of who ought to be categorised as `older'. Age is not merely a product of individual processes, as examined in Johfre and Saperstein's and our typologies, but is also influenced by interactional (how people respond to perceived age, e.g.\,with prejudice or not), organisational (ways of managing age groups, e.g.\,the pressure to develop technological solutions to a so-called `crisis of care') and cultural processes (notably, age stereotypes) \cite{johfre2023social}. Taken together, these factors and processes are inherently intersectional and play out across the lifecourse; and HCI research, as we have seen, is partly contributing to this intersection not only in the technology it offers `older adults'  but in the way it approaches them, i.e.\,in how it identifies and addresses their `problems'. Technology and its associated research and development communities, including CHI, are deeply implicated in the social construction of age at all of these levels in ways that deserve much deeper examination.

\section{Age as a system of (digital) stratification}
\label{inequality}

In the preceding sections, we adapted Johfre and Saperstein's sociological typology of multidimensional age by first updating operationalisation notes to speak to HCI examples and, second, exploring what this adapted typology can tell us about the measurement of age(ing) in HCI research. We noted  HCI research has been contributing to some of the constrained ways age and ageing might be understood by separating factors that might be better treated as connected and intersectional, and in this way has added to a view of ageing that oversimplifies.  Yet we also find when the HCI literature is examined holistically, it is rich in its attending to the multiple factors that constitute and shape what ageing might be.
    Given this, we now zoom out from the specifics of the \textit{typology} to consider how treating age as a multidimensional social construction provides a useful new \textit{framework} for understanding digital inequities.

To succinctly capture what Johfre and Saperstein explain at length, 
while age research (broadly speaking) typically examines individual age dimensions as predictors of various outcomes, a multidimensional view treats these dimensions as resources a person has for enacting age. Enacting age becomes important when age is used to sort people into categories which delimit the rights and responsibilities of the person. Differences in the societal expectations of older adults' technological engagement and proficiency---i.e.\,the well-known issue of the age-based digital divide\footnote{We adopt this phrase from \cite{neves2018old} to refer to the persistent finding that older adults lag behind younger adults in terms of technology uptake and use. This is evidenced in a vast body of survey-based research (e.g., \cite{OECD,Pew,LLoyds23}).}---is an example of how this construction is consequential and why it might matter for the field of HCI. In what follows, we (i) propose a new way of understanding the relationship between age (and associated measures) and the macro-phenomenon of the digital divide, (ii) discuss the implications of this relationship in terms of when and why this divide can be (but isn't inherently) inequitable, and (iii) explore opportunities for enriching HCI and Ageing discourse through this reconceptualisation of age and the sequelae of its construction.

The challenges of operationalising chronological age notwithstanding, substantively, one way to view ageing as a \textit{social process}  \cite{jones2023s,lazar2021adopting,vines2015age} is to consider how it operates as an axis where resources can be accumulated (or not) \cite{Dannefer03}. Time itself is a resource, and intangible, non-fungible resources such as technical skills (i.e., \textit{expertise}) and psychosocial skills (e.g.\,self-confidence, self-efficacy; a.k.a.,\textit{wisdom})  require time to develop; chronological ageing marks the passage of time during which this development could occur. 
Whether these resources are accumulated, decumulated, or static over the lifecourse, however, is not purely a function of time; otherwise, we would expect age to be the only predictor of these resources.  They could be thought the consequences of \textit{tenure},  to borrow the phraseology from above. An intersectional view,\footnote{For brevity, we assume readers are familiar with the concept of intersectionality. For more, see \cite{Davis_1983,Crenshaw_1991}.}  on the other hand, highlights the structural forces that shape whether and/or how resources are accumulated over the lifespan. 

One way of understanding older adults' relative lack of engagement with technology is as indicative of resource \textit{decumulation} over time, whether in terms of material resources (e.g.\,to purchase technology and supporting infrastructure), technical skills, and/or free time to invest in developing these skills, all of which being subject to broader structural forces. Ageing provides time for (i) consequential decumulating events to occur (e.g.\,illnesses, periods of caregiving, etc) and (ii) the inequalities found to be risk factors for digital exclusion at any age to amplify to significance, which may help explain the enduring predictiveness of age in digital exclusion \cite{HouseOfLords,friemel2016digital}. But aligning with the recommendation for HCI to critically engage with the concept of lifecourse \cite{vines2015age}, we suggest there is a need to more thoroughly track the kinds of life events that precipitate resource decumulation  to understand opportunities for early intervention for mitigating their cascading consequences for technology use in later life. Additionally, an intersectional lens would be particularly helpful for unpacking why resources decumulate or accumulate differentially, and for whom. Gender norms and childbearing/rearing expectations, as alluded to above, may be highly consequential for structural inequities across the lifecourse which result in digital inequity for one gender but not for another, as a case in point. We also note the work of social reproduction is both gendered and racialised \cite{ONS-SRT}, and is not only systematically devalued \cite{bhattacharya2017social, federici2019social,clair2021watershed} (with implications for material resource accumulation) but also offers fewer opportunities for technical skills accumulation for those undertaking it. Together, this suggests the need for much deeper engagement, too, with interactions between age, gender, and race---and other axes of stratification, such as disability and class---within HCI and Ageing.

Another interpretation of older adults' relative form of engagement (disengaged, say, or more engaged, depending on a relevant measure)  with technology is as an \textit{accumulation}. Accumulation can take many forms, aggregating into certain forms of privilege. Privilege, itself, also accumulates differentially---the passage of time being a necessary but not sufficient condition for the accumulation of some forms of privilege and not others. 
Notably, psychosocial skills can be essential for enacting agency and self-determination, including resistance to or engagement with digital technologies. Likewise, and in terms of \textit{family role}, having younger relatives that can act as technology mediators \cite{Tang2022} and are expected to look after elders in this way (a norm that is highly culturally dependent \cite{Gutierrez2016,Gutierrez17}), can be another privilege that diminishes an older adult's need to use digital technologies themselves. As a result, whether disengagement from technology is a consequence of privilege---a benefit of a kind---or a disadvantage needs care to specify. Non-adoption of technologies to support ageing in place could reflect accumulation not just of family labour resources (i.e., younger family members doing things for the ageing) but also material resources to pay for a private carer or residential care. (Intersectional structural forces can also differentially shape access to all of these resources across the lifecourse.) Our point is non-use of digital technology is sometimes, but not always, inequitable. Inequality of outcome---that is, limited or non-use of technologies by some people---is inequitable where there is a \textit{desire} to engage with these technologies, but a structural barrier limits the extent to which this desire to engage can be realised. Unarguably, there are structural barriers driving some limited/non-use of technology that require redress.
Where limited/non-use is a \textit{choice}, however, inequality of use is a reflection of agency being enacted and is not only \textit{not inherently inequitable}, but may in fact be indicative of later life ``success'' from the older adults' own perspective \cite{vines2015age}.

This insight has important implications for how we think about the consequences of \textit{stage in life transitions}, such as retirement. On the one hand, retirement is accompanied by a loss of on-the-job technology training that may precipitate decumulation of technical skills, and reliance on a pension can mark a sharp decumulation of material resources. On the other hand, whereas working life creates strong incentives to invest resources in the maintenance of specific technological skills, retirement means greater choice regarding the technologies one engages with. Even though the result might be the same, i.e.\,decumulation of technical skills, the consequence of that decumulation depends on how valuable the thing is which one is losing skills in. For example, people may happily trade proficiency with video conferencing tools for the freedom to never have to use them again. So it is not decumulation (nor accumulation) of skills itself that ought to concern HCI and Ageing, but rather whether a decumulation or an accumulation is material to people's agency and overall well-being in later life. This, of course, interacts with the extent to which these are \textit{structurally supported}---that is, whether the greater agency people may look forward to in their retirement is constrained through the embedding of technology in critical services. This suggests a need for deeper exploration of what opportunities older adults are being systematically excluded from by decumulation of digital skills or accumulation of other resources, which of these  negatively or positively impact their lives, and what can be done to mitigate critical skills loss and/or its relative impact. What is negative for one individual will be formed by their overall intersectional position, what is positive for another a different arrangement of intersecting factors.

Our final point builds on the insight that people perceive and perform their age in relation to various dimensions of age. As Johfre and Saperstein explain \cite{johfre2023social}, some of these dimensions are more constrained than others (people have more control over their \textit{cultural consumption}, for example, than their \textit{physical} age, and more control over their \textit{physical} age than their \textit{chronological}  age), but also one's situatedness within different systems of stratification determines the particular constraints they encounter for these different dimensions. Just as some people may have the material resources for cosmetic surgery to alter perception of age through physical cues, some people (and not others) will have resources to purchase an array of technologies that support healthy ageing / ageing in place, along with the requisite foundational skills (and absent any significant disabilities) enabling them to use these tools. We note that the strong bias in HCI and Ageing towards samples of `white', middle-class people based in Global Minority countries in the corpus, while convenient for recruitment, contributes to the construction of a very narrow, and privileged, view of age(ing) (see also \cite{burghardt2022age}), and restricts consideration of how the passage of time is experienced intersectionally, how this matters for (non-)engagement with technology, and how (non-)engagement interacts with disparate outcomes such as life expectancy. CHI needs to look beyond its historical roots in particular cultures and places if it wants to understand ageing in the round. But also, because use of technologies is one of the more ``active-performance aspects of age'' \cite{johfre2023social} (\textit{cultural consumption}, again, being a dimension that has the potential to be less constrained than other dimensions), HCI and Ageing research that critically examines the reasons for (non-)engagement with technology has the potential to make an important discipline-spanning contribution to the wider literature on inequality at an international level. While reasons for non-use are frequently explored in the CHI literature, i.e.\,through participants' accounts, there is opportunity for much more comprehensive exploration of how various other age dimensions interact with \textit{cultural consumption}. We propose that this framework could be used to scaffold data collection (e.g.\,interviews, surveys) in ways that elicit introspection on how one's individual/intersectional, multidimensional experience of age relates to their patterns of dis-/engagement with technology. This more methodical approach would enable the field to advance theorisation on the significance of age to HCI.

Taken together, we argue the `age-based digital divide' may well be real---which is to say, ageing as a physiological \textit{and social} process may contribute to recognisable patterns of technology use and an overall reduction in technology engagement, as the research in and beyond HCI repeatedly finds---but if so it is because of multiple, intersecting and sometimes diverging factors. Hence the term can be troublesome, as it suggests something simple and uniform, not a complex set of entangled issues giving rise to subtly different presentations of superficially similar phenomena bound up with all too often unexamined premises about age and ageing in the research. Yet this has not negated the uncovering  of multiple factors of concern in the literature. Despite its noted limitations, HCI and Ageing shows a remarkable corpus of findings that provides a strong basis for undertaking a deeper analysis of these matters.

\section{Conclusion}

In this paper we have organised CHI's vast and varied literature on ageing into a coherent framework that can scaffold substantive and productive engagement with HCI and Ageing's core concepts of `age' and `ageing'. 
Appreciating how age is understood, interpreted, and constructed in HCI enables the field to re-examine what has been taken for granted about the relationship between age and observed differences between categories of `old' versus `young' users, including differences in rates of technology adoption and types of use. We have highlighted not only the pressing need to interrogate why time since birth is so important in HCI, but also how it links to resource decumulation. Moreover, our intersectional lens on ageing has highlighted that most studies focus on a Global Minority population. These insights represent substantial gaps in extant literature and point to an important agenda for future research. 

We have argued a narrow focus on naturalised age (i.e.\,age being conceptualised as an objective function of time) and reliance on the poor proxy of chronological age within extant HCI and Ageing research has obfuscated the role of intersectional dis/advantage, multiplied across the lifecourse, in influencing both the technological and material resources that determine whether one has the means to interact successfully with, and even dis/engage with, technology in later life. Our adapted framework of age in HCI can help researchers dismantle the monolith of `older adults' and inspire more deliberate examination of the dimensions of age that matter to a given context or research question. We propose this puts the field in a better position to explore how \textit{multiple} dimensions of age combine in ways that prove salient to a person's ability and desire to engage with technology; and how such combinations reflect that person's intersectional position in society, enabling a more nuanced discourse on age-patterned digital inequality vs.\,inequity.


\begin{acks}
This work is supported by the EPSRC funded grant Equity for the Older: Beyond Digital Access (EP/W025337/1).
\end{acks}

\bibliographystyle{ACM-Reference-Format}
\bibliography{bibliography} 

\section*{Appendix}

Additional details of the corpus creation are included below.

\subsection*{Corpus A} A full-text search was conducted of the ACM Digital Library using the advanced search function to apply the filters `CHI: Conference On Human Factors In Computing Systems' and `January 2014 -- June 2024'\footnote{Search date: 20 June 2024. This date range and filter includes exactly 10 years of the CHI conference, but excludes TOCHI papers presented at the conference.} and including ``older adults'' anywhere in the text. This returned 804 results, which were refined to exclude Extended Abstracts for courses, workshops, panels, SIGs, town halls, session descriptions, and lists of proceedings (n=72). Duplicates otherwise meeting inclusion criteria were removed (n=2); and publications from other conferences incorrectly linked to CHI were also removed (n=6). The remainder were manually screened for relevance. 

Publications using ``older adults'' in the title and/or abstract were included; so, too, were publications using ``older people'', ``elderly'', and ``senior'' (if determined to mean ``senior citizen'') in the title and/or abstract. Publications not using any of these terms in their title or abstract were included if they substantively focused on older adults based on the abstract (e.g.\,if talking about retired individuals or grandparents). In cases where this could not be determined from the abstract, the PDF of the publication was searched for ``older adults'' to understand the context of use of this phrase. Where ``older adults'' was used as a keyword, the publication was included. Where the publication mentioned challenges, opportunities, behaviours and/or attitudes of older adults anywhere in the text, the publication was included; as were publications reporting differences found between ``older'' and ``younger'' adults, if exploring multiple age groups. Publications were excluded if ``older adults'' occurred only in the list of references,\footnote{Papers about `older adults' were often cited 
because they provided transferable insights for a given research context. As an example, a CHI paper entitled, ``Examining Crowd Work and Gig Work Through The Historical Lens of Piecework'' \cite{alkhatib2017examining} was returned in our search---not because it focuses on older adults, but because in exploring crowd work it cites a paper that is in our corpus \cite{Brewer16why} that examines crowd work by older adults.}  or only in reporting findings of other studies (e.g.\,in the background section). Publications were also excluded if specifically focusing on caregivers (n=17) or health practitioners (n=1), unless the caregivers themselves were older adults (n=1) \cite{Tixier16}.

Decisions about inclusion were more difficult for publications on the topics of cognitive decline (e.g.\,dementia, Alzheimer's) and physical decline, disease, and/or disability (e.g.\,stroke, cancer, Parkinsons, COPD, deafness, blindness). Many papers on dementia specifically avoided classing it as an `older adult' disease, as it has been known to affect people as young as 30, depending on the type (see \cite{ma2024bridging}). Publications on cognitive decline and/or dementia were included in Corpus A if authors explicitly focused on individuals aged 65+ and/or framed cognitive decline as a function of age; and publications on physical decline, disease and/or disability were excluded unless authors made an explicit connection to advanced age. The resulting corpus was comprised of 234 publications.

\subsection*{Corpus B} 
A separate ACM Digital Library search was conducted applying the filters `CHI: Conference On Human Factors In Computing Systems' and `January 2014 -- June 2024', but this time including ``older adults'' OR ``older people'' OR ``senior citizen'' OR ``elderly'' in the abstract. This returned 222 results, which were refined to exclude Extended Abstracts for courses, workshops, panels, SIGs, town halls, session descriptions, and lists of proceedings (n=19) and incorrectly linked publications (n=4), as per protocol for Corpus A. The remaining 199 publications were manually screened for relevance. 

Consistent with the exclusion criteria for Corpus A, publications were excluded if not exploring challenges, opportunities, behaviours and/or attitudes of older adults (n=4); or if specifically focusing on caregivers (n=3) or family of the older adult (n=1), unless the caregivers themselves were older adults (n=0). The resulting corpus was comprised of 191 publications.

\subsection*{Final corpus}
Corpus A and B overlapped significantly, as one would expect: 

\begin{figure}[h]
  \includegraphics[width=.4\textwidth]{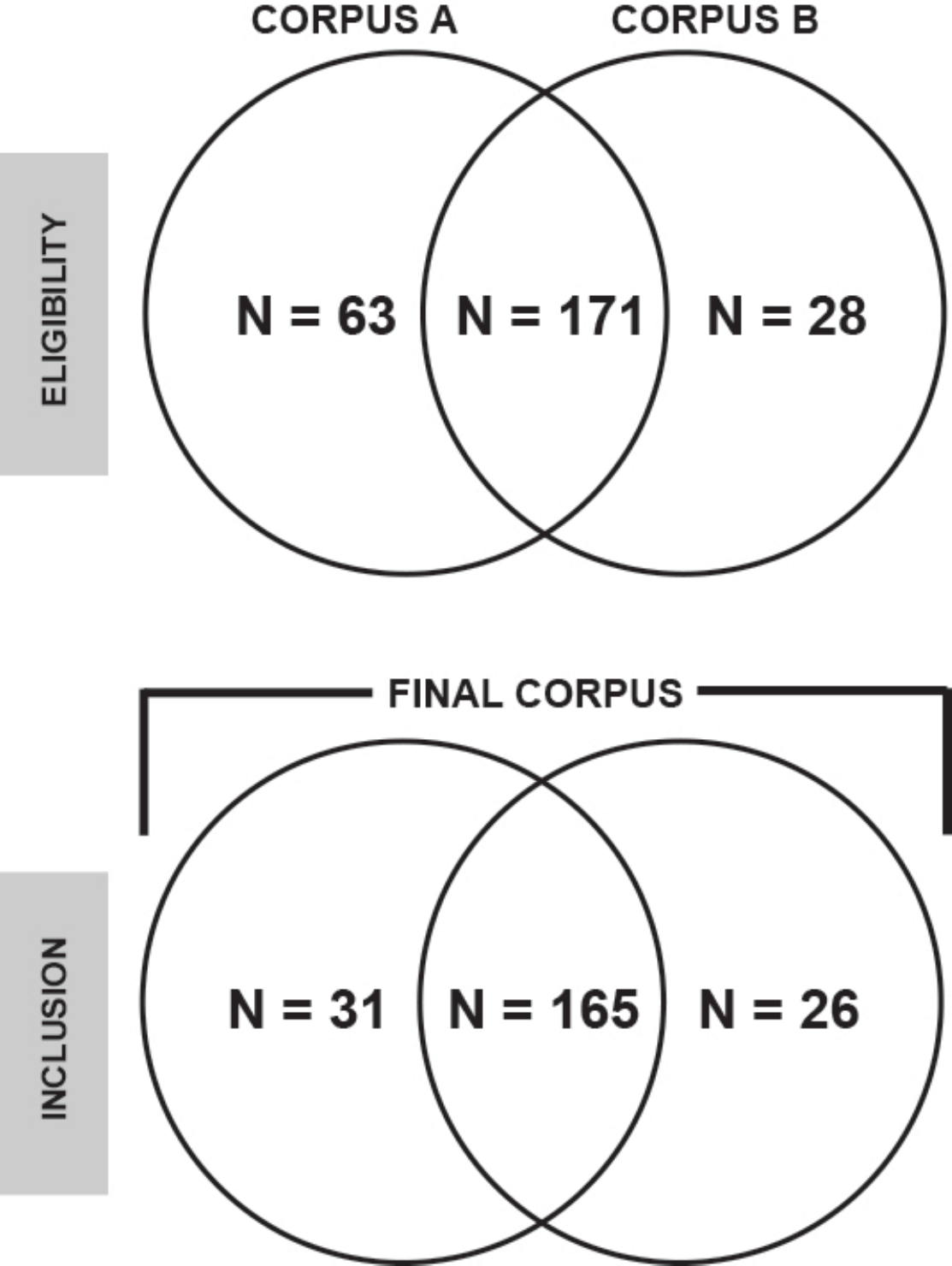}
  \caption{Venn diagram of publications from Corpus A and Corpus B.}
  \label{fig:Venn}
\end{figure}

To ensure inclusion of all papers motivated by and/or making a contribution pertaining to understandings of `older adults', the 63 publications from Corpus A that did not overlap with Corpus B (i.e.\,that discussed ``older adults'' somewhere in the paper but not in the abstract) were manually screened for relevance. 

Publications with ``older adults'' (or variants included in the search string for Corpus B) in the title and/or author keywords were included. For the remainder, where the title and/or keywords plausibly pertaining to `older adults' (e.g.\,ageing, elders, grandparent, retirement), the introduction section was read to determine relevance. Publications about `grandparents' were only included if they focused on something particular to `older adults', as opposed to merely exploring inter-family relationships. Two (both) papers exploring menopause \cite{Burst24,New21}\footnote{In support of this basis for exclusion, \cite{Burst24} explicitly warn technology designers of the perils of viewing menopause apps as being ``just for older adults''.} and 1 paper exploring ``older job seekers'' \cite{10.1145/3613904.3642959} were excluded because they used `older' comparatively within a category of younger individuals, i.e.\,were looking at middle-age. (Note: we differ from \cite{petrie2023talking} in making this exclusion decision.) A total of 31 papers were found to be relevant, and were thus added to the 191 papers in Corpus B. The final corpus was comprised of 222 publications. This includes research articles (i.e.\,full papers; n=126) and Extended Abstracts for: Late Breaking Work and Work-In-Progress (n=39); alt.CHI (n=3); Demo and Interactivity (n=7); Case Study (n=3); Doctoral Consortium (n=5); Student Research Competition, Student Design Competition, and Student Game Competition (n=8). The remaining publications are Extended Abstracts without a conference track label (n=31). 

\newpage

\subsection*{Tabular summary of results}
\begin{table*}[h]
    \begin{tabular}{|p{.13\linewidth}|p{.27\linewidth}|p{.45\linewidth}|p{.05\linewidth}|}
        \hline
   \textbf{Dimension}    & \textbf{Example measures} & \textbf{Operationalisation in HCI \& Ageing}   & \textbf{\%} \\ \hline \hline
        Chronological                                                                                           & Time since birth  & Calculated via birthdate/year of birth, or self-reported age of participants  & 84.2\%
        \\ \hline
         {}                                                                                           & Life expectancy at age X 
  & Years left to live and use technologies, calculated based on birth year/cohort relative to shifting demographics & 10.8\%
  \\ \hline
           {}                                                                                           & Healthy life expectancy at age X 
 & Ability to age healthily, independently, and in one's own home & 35.1\%
  \\ \hline
             Generation                                                                                           & Generation/cohort
& Birth cohort calculated via chronological age bands, shared socio-historical background, or shared exposure to technologies; relative inter-family position & 13.1\%
  \\ \hline 
               {}                                                                                                                                                                                 & Institutional cohort 
& Category within society entitled to social protection and/or requiring management, e.g., vulnerable, needing care;  category within digital economy, e.g., historically marginalised & 23.9\%
  \\ \hline
                 Physical                                                                                          & Biomarkers
& Gait/stride length, skin elasticity, pupil dilation & 1.8\%
  \\ \hline
                   {}                                                                                          & Health and energy
& Functional abilities (e.g., eyesight, hearing, motor skills, haptic control, balance, swallowing, sensory perception); overall health (e.g., disease prevalence, severity, or co-morbidity); muscle strength and fitness; fatigability & 24.8\%
  \\ \hline
                   {}                                                                                          & Appearance and embodiment
& Voice quality, ``voice age''; speech features; movement repertoire and slowness & 2.7\%
  \\ \hline
                     Psychological                                                                                          & Cognitive ability
& Processing speed, spatial cognition, cognitive map decay; memory (working, procedural, long-term), input stumble; intelligence (fluid and crystallised) and functional neuroplasticity; social functioning and communication; decision making, multitasking, attentional control/splitting, cognitive inhibition; neurocognitive disorder (mild cognitive impairment and dementia) & 27.9\%
  \\ \hline
                       {}                                                                                          &  Emotional maturity
& Measured in rates of technology adoption and interaction behaviours: confidence/hesitance; fearfulness; circumspection; autonomy, collaborativeness; creativity; adventurousness; willingness to learn
  & 4.5\%
 \\ \hline
Life stage                                                                                          &  Age category
& ``Identity age'' (fluid and actively negotiated, e.g., situated elderliness); negative self-stereotyping (self-beliefs about ICT skills); negative other-stereotyping (distancing oneself from `older adult' category)  & 2.7\% 
  \\ \hline
  {}                                                                 &  Stage in life  transitions
& Stages include retirement (financial stability), social isolation (living alone, distanced family), restricted mobility / disability, (pseudo-) institutionalisation & 31.1\%
  \\ \hline
  Responsibility                                                                 &  Family role
  & Social role as parent, grandparent, spouse; role as carer or cared-for individual; role within intra-family/household technology relations (receiver of IT support, sharer of data/passwords) & 8.1\%
  \\ \hline
    \end{tabular}
    \caption{\label{tab:table-results1} HCI and Ageing typology of age: coding results.}
\end{table*}


\begin{table*}[h]
    \begin{tabular}{|p{.13\linewidth}|p{.27\linewidth}|p{.45\linewidth}|p{.05\linewidth}|}
        \hline
   \textbf{Dimension}    & \textbf{Example measures} & \textbf{Operationalisation in HCI \& Ageing}   & \textbf{\%} \\ \hline \hline
             {}                                                                 &  Civic rights and responsibilities
& Whether the person can still drive and contribute to healthcare decisions; contributing to society and community (e.g., inter-generational mentorship, co-design, content creation) & 5.9\%
  \\ \hline
  {}                                                                 &  Workplace responsibility
& Whether the person is still in/seeking employment (e.g., need of technology upskilling) & 1.4\%
%
  \\ \hline
  Experience                                                                 &  Tenure
& Time spent using technologies (e.g., when they first learned); time spent using newer technologies (e.g., familiarity) & 5.9\%
  \\ \hline
            {}                                                                 &  Expertise
& Technology expertise (e.g., comfort levels using technology, susceptibility to digital harms) & 14.9\%
  \\ \hline
              {}                                                                 &  Wisdom
& Competence and self-efficacy; attitudes and integrity (e.g., asserting one's values in adopting or rejecting technology) & 11.3\%
  \\ \hline
                Cultural consumption                                                                 &  Aesthetic preferences
& Preferences in technology adoption and use; preferences in robot/chatbot/VA appearance and behaviour, icon style, or tacility  
  & 9.0\%
  \\ \hline 

                   
                  {}                                                                 &  Activities and interests
& Behaviours (e.g., frequency of exercise), orientations (what is considered meaningful and worth doing), and hobbies (e.g., crafting/making, gaming, crowdworking, making music, volunteering, lifelong learning)
  & 16.7\%
  \\ \hline
Other signifiers                                                                 &  First names
& Age appearance of avatar & 0.5\%
  \\ \hline
    \end{tabular}
    \caption{\label{tab:table-results2} HCI and Ageing typology of age: coding results (continued).}
\end{table*}

\end{document}


\title{Not Just a Number: A Multidimensional Approach to Ageing in HCI}

\renewcommand{\shortauthors}{Knowles, et al. \copyright January 2025}


\nocite{*} 
\renewcommand{\refname}{Supplemental Bibliography}
\bibliographystyle{ACM-Reference-Format}
\bibliography{supplementary}